\documentclass[a4paper,fleqn]{cas-dc}

\usepackage[numbers]{natbib}
\usepackage{booktabs}

\graphicspath{{./}}

\begin{document}

\shorttitle{Adjoint optimization for surgery of septal deviations}
\shortauthors{Macellari, Schillaci, Tanzini, Trimarchi, Quadrio}
\title [mode = title]{An adjoint-based approach for the surgical correction of nasal septal deviations}

\author[1]{Marcello Macellari}
\author[1]{Andrea   Schillaci}
\author[2,3]{Umberto Tanzini}
\author[4]{Matteo    Trimarchi}
\author[1]{Maurizio Quadrio}[orcid=0000-0002-7662-3576]
\cormark[1]
\ead{maurizio.quadrio@polimi.it}
\affiliation[1]{Dept. of Aerospace Science and Technologies, Politecnico di Milano, Campus Bovisa, Milano 20156, Italy}
\affiliation[2]{Division of Head and Neck, Otorhinolaryngology unit, IRCCS San Raffaele Scientific Institute, Milano, Italy}
\affiliation[3]{School of Medicine, Vita-Salute San Raffaele University, Milano, Italy}
\affiliation[4]{Department of Otolaryngology -- Head and Neck Surgery, Ente Ospedaliero Cantonale, Ospedale Regionale di Lugano -- Universita' della Svizzera Italiana, Lugano, Switzerland}
\cortext[1]{Corresponding author}

\begin{abstract}
Deviations of the septal wall are widespread anatomic anomalies of the human nose; they vary significantly in shape and location, and often cause the obstruction of the nasal airways. 
When severe, septal deviations need to be surgically corrected by ear-nose-throat (ENT) specialists.
Septoplasty, however, has a low success rate, owing to the lack of suitable standardized clinical tools for assessing type and severity of obstructions, and for surgery planning.
Moreover, the restoration of a perfectly straight septal wall is often impossible and possibly unnecessary. 
This paper introduces a procedure, based on advanced patient-specific Computational Fluid Dynamics (CFD) simulations, to support ENT surgeons in septoplasty planning. 
The method hinges upon the theory of adjoint-based optimization, and minimizes a cost function that indirectly accounts for viscous losses. A sensitivity map is computed on the mucosal wall to provide the surgeon with a simple quantification of how much tissue removal at each location would contribute to easing the obstruction. 
The optimization procedure is applied to three representative nasal anatomies, reconstructed from CT scans of patients affected by complex septal deviations. 
The computed sensitivity consistently identifies all the anomalies correctly. 
Virtual surgery, i.e. morphing of the anatomies according to the computed sensitivity, confirms that the characteristics of the nasal airflow improve significantly after small anatomy changes derived from adjoint-based optimization.
\end{abstract}




\maketitle

\section{Introduction}

Nasal Airway Obstructions (NAO) are one of the main medical conditions for which patients consult ear-nose-throat (ENT) doctors \cite{chandra-etal-2009}. 
Indeed, more than one third of the world's population is affected by NAO \cite{jessen-malm-1997, li-etal-2020}, with a reduced nasal airflow that impacts the quality of life \cite{rhee-etal-2003, udaka-etal-2006}. 
Among the main aetiological factors causing NAO, septal deviations are extremely common, with a prevalence of 76\% \cite{clark-etal-2018}. 
In severe cases ENT specialists resort to surgical corrections via septoplasty to restore a functional respiration.
Because of a compensation mechanism, septal deviations often induce an hypertrophy of the contralateral turbinate \cite{kang-2022}, which potentially leads to obstruction after the septum is surgically straightened.
Even though severe complications are rare and arise in the 3\% of cases only \cite{dabrowska-bien-etal-2018}, once the actual post-surgical benefits are considered, the success rate of septoplasty falls below 50\% \cite{tsang-etal-2018} or even less \cite{sundh-sunnergren-2015}, resulting in a relevant social and financial burden for the healthcare system. 
This state of affairs can be ascribed to the lack of standardized and reliable clinical tools to assess type, severity and consequences of obstructions, which would provide surgeons with essential information when deciding on the most appropriate surgical action \cite{dinis-haider-2002,roblin-eccles-2002}. 

To date, no universally accepted criteria exist for selecting patients for septoplasty, not to mention for a detailed surgery design. Objective indicators have been proposed to select patients for surgery \cite{holmstrom-2010}, but the limited correlation between clinical measurements (rhinomanometry, acoustic rhinometry) and the patients' subjective perception of the quality of the airflow \cite{andre-etal-2009} have prevented so far the establishment of objective criteria that are generally accepted and used.

Computational Fluid Dynamics (CFD) is nowadays recognized as a valuable tool to study the nasal airflow, to quantify its characteristics and to relate them with the patient's anatomy \cite{kleinstreuer-zhang-2010, calmet-etal-2021}. 
Techniques range from relatively inexpensive simulations that take advantage of turbulence modelling \cite{li-etal-2017} to larger-scale and higher-fidelity calculations \cite{calmet-etal-2016}. Cases with NAO are also considered \cite{leong-etal-2010}. 
For instance, Refs. \cite{chen-etal-2009}, \cite{liu-etal-2012}, and \cite{janovic-etal-2020} used CFD to evaluate changes in the respiratory pattern when septal deviations of different type and severity occur. Refs. \cite{malik-etal-2021} and \cite{radulesco-etal-2019}, instead, performed simulations of the nasal airflow to analyze what quantities, among those measured with existing clinical tools, are the most representative of the perception of patients. 
Several studies directly examined the effect of septoplasties. In particular, Campbell et al. \cite{campbell-etal-2021} focused on anterior septoplasty, and studied the CFD-computed nasal airflow of ten healthy anatomies modified to introduce NAO, to understand how surgically widening the region with minimal cross-sectional area correlates with the actual benefits of surgeries. 
Ramanthan et al. \cite{ramanathan-etal-2021}, on the basis of CFD results for twelve pre- and post-operative CT scans, suggested the main regions of obstruction to be often identified by locally high pressure, velocity and shear stress. 
However, even after the obstructive region has been identified, the available studies are unable to specify in which way the area of a certain cross-section should be enlarged.

All the CFD works addressing septoplasty planning rely, to a varying extent, on a subjective interpretation of the CFD results. 
While a strong local minimum of the cross-sectional area is certainly bound to determine an obstruction, associating NAO (and the corresponding strategy for its surgical correction) only to the minimal cross-sectional area, or to the local extremum of CFD-computed quantities, may be too simplistic, as it neglects the anatomical complexity of the cross-section itself and the non-locality of the fluid flow equations: the flow field observed in one specific location is certainly affected by the anatomy elsewhere. 
Moreover, the value of the minimum cross-sectional area does not inform the surgeon on the directional changes of the velocity vector induced by enlarging different portions of the minimal-area section. A certain enlargement might be achieved by acting on different portions of the minimal-area section, with significantly different consequences on the airflow.

The goal of this work is to introduce a novel CFD-based procedure for  patient-specific septoplasty planning. 
The approach includes elements from the adjoint-based theory of shape optimization, that was introduced in fluid mechanics 50 years ago, see e.g. Ref.\cite{pironneau-1974}, and is becoming progressively popular in CFD. 
The procedure consists of two steps. In the first, the airflow within the nasal cavities is simulated via conventional CFD. The second step then uses the computed flow field to solve an optimization problem, by finding the minimum of a cost function.
The end result can be put in the form of a sensitivity map, i.e. a distribution on the whole nasal surface of a scalar quantity that, at each point, quantifies to what extent the displacement of said point, as a consequence of a surgical action, is favorable or counterproductive in terms of minimization of the cost function.
In this paper, the adjoint-based method is presented and exemplified via application to three nasal anatomies of patients affected by complex septal deviations. 
An in-depth clinical analysis by ENT surgeons confirms the surgical feasibility of the indications derived from the sensitivity. Virtual surgery is also carried out, to confirm that the computed sensitivity leads to significantly improved nasal resistance with minimally invasive surgeries.

The paper is structured as follows. A brief, non-technical summary of the adjoint-based optimization is given in Sec.\ref{sec:theory}, followed by Sec.\ref{sec:methods} which contains a description of the numerical methods employed in the present work, including the adaptation of the theory to the specific problem. Results for the three patients are presented in Sec.\ref{sec:results}. An in-depth discussion, which includes comments on the essential aspects of the procedure as well as the virtual surgery study, is provided in Sec.\ref{sec:discussion}, and Sec.\ref{sec:conclusions} is devoted to conclusions.  

\section{Adjoint-based optimization}
\label{sec:theory}

Central to the present work is the adjoint optimization technique used for computing the surface sensitivity in a shape optimization problem. 
The main theoretical and technical aspects behind the adjoint formulation are briefly summarized below, while the tailoring of the method to septal deviations is described in Sec.\ref{sec:methods}. 
The interested reader is referred to recent review papers \cite{dilgen-etal-2018, alexandersen-andreasen-2020} for additional information on adjoint optimization. 

In general, in shape optimization one seeks the optimal shape of a two- or three-dimensional object (defined through a cloud of points) that minimizes a certain (known) cost function while satisfying a set of constraints. 
In the CFD context, and in a formulation that naturally provides results defined on the original shape, this can be expressed as:  
\begin{equation}
\begin{split}
\mbox{minimize} \;f=f(\mathbf{U},\beta) \\ 	\mbox{subject to} \;R(\mathbf{U},\beta)=0
\end{split}
\label{eq:opt}
\end{equation}
where $\beta$ is the set of control variables (e.g. the wall-normal displacements of the nodal points of the surface) which define a change in shape, and $f$ is the cost function to be minimized by properly choosing $\beta$. $\mathbf{U}$ is the set of flow variables (e.g. velocity and pressure), and $R$ a set of constraints, indicating that values of the flow variables are not arbitrary, but must obey the differential equations governing the fluid flow. 
Let us consider the steady incompressible Navier--Stokes equations:

\begin{equation}
\begin{cases}
\nabla \cdot \mathbf{{u}} = 0 \\
\left( \mathbf{u} \cdot \nabla \right ) \mathbf{u}= - \nabla p +
\nu \nabla^2 \mathbf{u}
\end{cases}
\label{eq:direct}
\end{equation}
where $\mathbf{u}$ is the velocity vector, $p$ the pressure divided by the (constant) density, and $\nu$ the kinematic viscosity of the fluid.
The optimization computes the sensitivity derivatives, i.e. the gradients of the cost function with respect to the control variables. The sensitivity describes how $\beta$ (a change in shape) affects $f$ (the cost function). 
The visualization of $\beta$, defined on the boundary only, highlights at a glance where altering the original shape is most effective at minimizing the cost function.

Adjoint optimization excels when a simple cost function is available, and the number of control variables is large (as in the present case). Indeed, the surface sensitivity is obtained at a computational cost that is independent upon the number of control variables: the procedure requires solving two systems of Partial Differential Equations (PDEs), whose computational cost is independent upon the number of elements of $\beta$. The first system is made by the usual governing equations of fluid dynamics, i.e. \eqref{eq:direct}, whereas the second one, approximately of the same cost, includes the so-called adjoint equations, which are derived from the Navier--Stokes equations. 

In practice, two approaches can be followed to derive the adjoint equations: a discrete one, known as "discretize then derive" and a continuous one, known as "derive then discretize" \cite{bewley-2001}. 
In this contribution, a continuous formulation specialized for internal flows and described by Othmer \cite{othmer-2008} is used as the starting point for the main analytical derivations. 
As often assumed in shape optimization problems, in this work the cost function includes contributions only from the boundary $\Gamma$ of the flow domain $\Omega$.

The optimization problem \eqref{eq:opt} is first reformulated in terms of Lagrange multipliers to account for the set of constraints $R$. The adjoint variables are introduced: the adjoint pressure $q$ (a scalar quantity) and the adjoint velocity $\mathbf{v}$ (a vector).
They have the same physical dimensions of their physical counterparts $p$ and \textbf{u}, but carry a different meaning. The resulting Lagrangian function, defined over the entire domain $\Omega$ reads:
\begin{equation}
L = f + \int \left( \mathbf{v},q \right) R \ d\Omega \ .
\end{equation}

Once $L$ is defined, the sensitivities can be computed by starting from the total variation of $L$, written as:
\begin{equation}
\delta L=\delta_{\beta}L + \delta_{\mathbf{u}}L + \delta_p L
\end{equation}
by separating the variations $\delta_{\beta}$, $\delta_{\mathbf{u}}$ and $\delta_p$. 

The adjoint variables, which can take arbitrary values, are chosen such that the sum of the variations of $L$ with respect to the state variables is null:
\begin{equation}
\delta_{\mathbf{u}} L + \delta_p L=0
\label{variation}
\end{equation}

At this point, the influence of $\beta$ on the Navier--Stokes equations can be evaluated. Indeed, the previous equality leaves $\delta_{\beta}L$ as the only contribution to the variation of the Lagrangian function. The final expression for the sensitivities is obtained as \cite{soto-lohner-2004}:
\begin{equation}
\frac{\partial f}{\partial \beta_i} \propto \mathbf{v}^i\cdot\mathbf{u}^i
\label{eq:sensitivity}
\end{equation}
where $\beta_i$ is the displacement, in the direction normal to the surface, of every point $i$ on the boundary, and $\mathbf{u}$ and $\mathbf{v}$ are the direct/adjoint velocities. 
It becomes evident that the surface sensitivity embeds information from the velocity field $\mathbf{u}$ (obtained via the usual CFD method), and from the adjoint field $\mathbf{v}$ (computed by solving the adjoint system of PDEs).

The adjoint system is derived from Equation \eqref{variation} by taking the required derivatives. For the present formulation, the resulting adjoint equations are:

\begin{equation}
\begin{cases}
\nabla \cdot \mathbf{v} = 0 \\
-\nabla \mathbf{v} \cdot \mathbf{u} - \left( \mathbf{u} \cdot \nabla \right) \mathbf{v} = -\nabla q + \nu \nabla^2 \mathbf{u}
\end{cases}
\label{eq:adjoint}
\end{equation}
in which the adjoint variables $\mathbf{v}$ and $q$ are the unknown to be computed, whereas the flow variables $\mathbf{u}$ and $p$ are regarded as known, as they have been previously computed by solving the direct system \eqref{eq:direct}. The adjoint equations are linear, and the adjoint velocity field is solenoidal.

Since the (still unspecified) cost function is assumed to contain contributions from the domain boundary only, the adjoint system \eqref{eq:adjoint} does not depend on the boundary shape, and enjoys general validity. Details on its derivation can be found in Ref.\cite{othmer-2008}. 

To solve the adjoint system, boundary conditions for the adjoint variables need to be specified. They are obtained by considering the contributions on the boundary $\Gamma$ that are present in Equation \eqref{variation} when the derivatives of the cost function are explicitly computed. Their definition depends on the specific cost function.

\section{Methods}
\label{sec:methods}

\subsection{Anatomies and discretization}

\begin{figure*}[ht]%
\centering
\includegraphics[width=\textwidth]{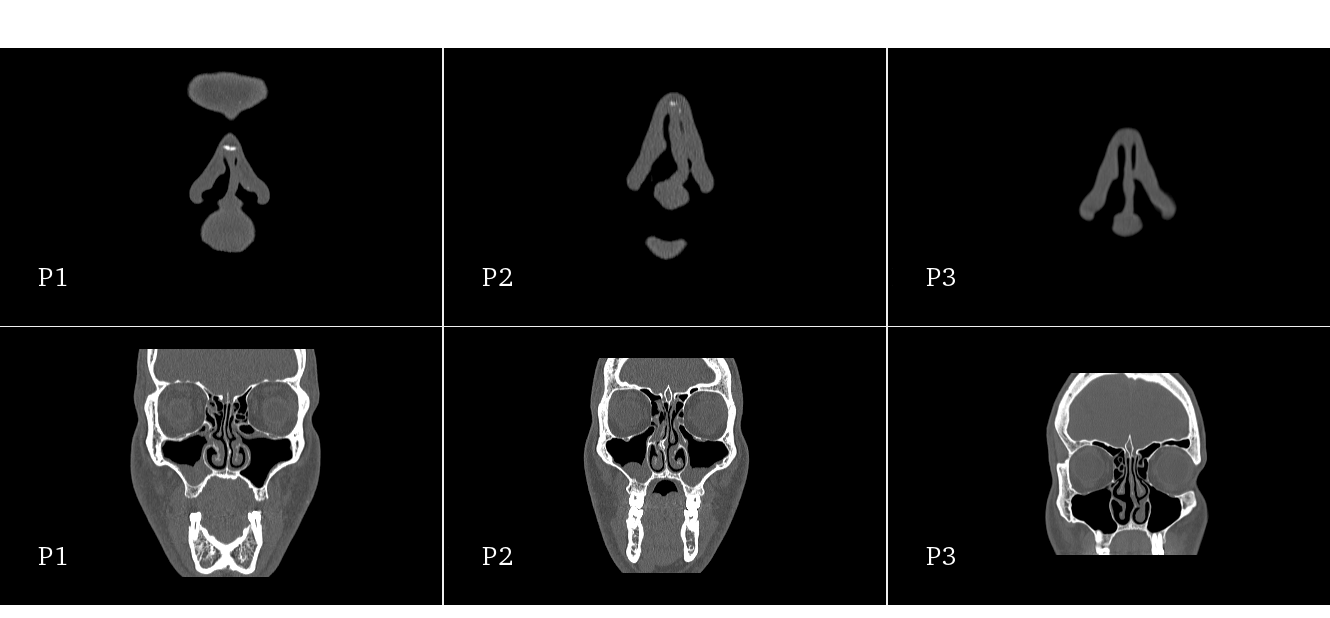}
\caption{Coronal sections of the three patients, visualized in correspondence of the most significant septal deviations in the anterior (top row) and posterior (bottom row) regions.}
\label{fig:anatomies}
\end{figure*}

In this study, three CT scans were selected from a pool of cases with complex septal deviations. They are shown in figure \ref{fig:anatomies}. The scans were extracted from a larger library of cases with septal deviations; they were selected by ENT surgeons among those providing average scan quality, and identified as cases where the surgical correction is not obvious in terms of its position and extent.

The scans, provided by the San Raffaele University Hospital, were obtained with a standard radiological protocol through a CT scanner with an acquisition matrix of $512^2$ pixels. The scanner is a GE Revolution Evo, with 128 slices. For the three patients, referred to as P1, P2 and P3, the spatial resolution of the scans in the sagittal-coronal plane is $0.39mm \times 0.39mm$, $0.31mm \times 0.31mm$ and $0.46mm \times 0.46mm$, and the gap between consecutive axial slices is $0.925mm$, $0.925mm$ and $0.4mm$ 

Patient P1, a 44-year-old caucasian male, presents a complex pattern of septal deviations, with a major left deviation of the quadrangular cartilage, and a small posterior deviation of the vomer bone. (As in the rest of the paper, the description adopts the point of view of the patient: hence, left/right should be intended as the patient's left/right.) 
It also presents a slight right deviation of the antero-superior portion of the nasal septum, at the articulation between the quadrangular cartilage and the perpendicular plate of the ethmoid bone. 
Patient P2, a 30-year-old caucasian male, has the anterior portion of the quadrangular cartilage deviated towards the left nasal fossa with a partial occlusion. Posteriorly, it presents an important right bone spur, bridging the middle meatus. Further posteriorly, another left bone spur reaches the posterior portion of the middle turbinate. 
Patient P3, a 40 year-old caucasian male, presents a nasal valve collapse (more evident in the left nostril) and a left septal deviation. The deviation involves the quadrangular cartilage, which is dislocated laterally and reaches the left inferior turbinate. Posteriorly it presents a left condro-vomerian spur that reaches the middle meatus. There is also a minor bony spur in the right posterior nasal fossa. Patient P3 probably underwent turbinoplasty before the CT scan, as the mucosa of the right inferior turbinate is less hypertrophic than normally expected. 

\begin{figure}[ht]%
\centering
\includegraphics[width=\columnwidth]{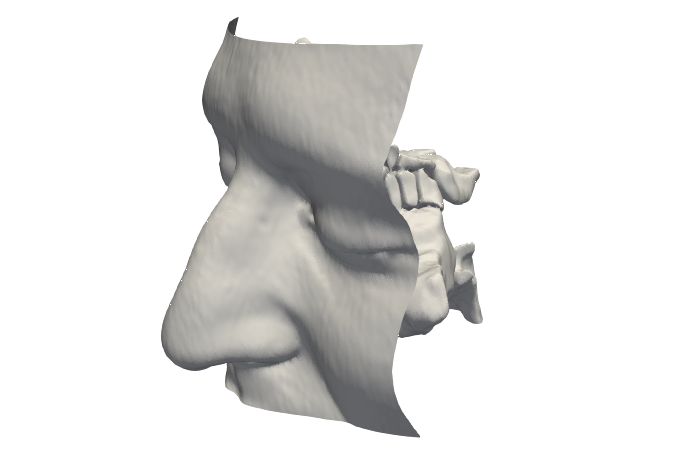}
\caption{Three-dimensional geometric model for patient P2.}
\label{fig:P2-STL}
\end{figure}

The selected scans are segmented with the free and open-source software 3D Slicer \cite{fedorov-etal-2012}. The three-dimensional reconstruction of the boundaries of the nasal airways is obtained by applying a segmentation threshold of -475 Hounsfield units, in accordance with results by \cite{nakano-etal-2013} and \cite{zwicker-etal-2018}. 
Figure \ref{fig:P2-STL} shows the final model for patient P2: the nasal airways up to the initial part of the nasopharynx and all paranasal sinuses are included.
The reconstructed geometries are used as input to create a computational mesh suitable for finite-volumes discretization. In this process, a spherical volume, with a diameter of $70 mm$, is placed around the nostrils to account for the external environment. As discussed in previous work \cite{covello-etal-2018}, such spherical volume places the actual inflow (where boundary conditions will be applied) far enough from the critical region of the nostrils, while keeping the total volume (and thus the computational cost) under control.

\begin{figure}[ht]%
\centering
\includegraphics[width=\columnwidth]{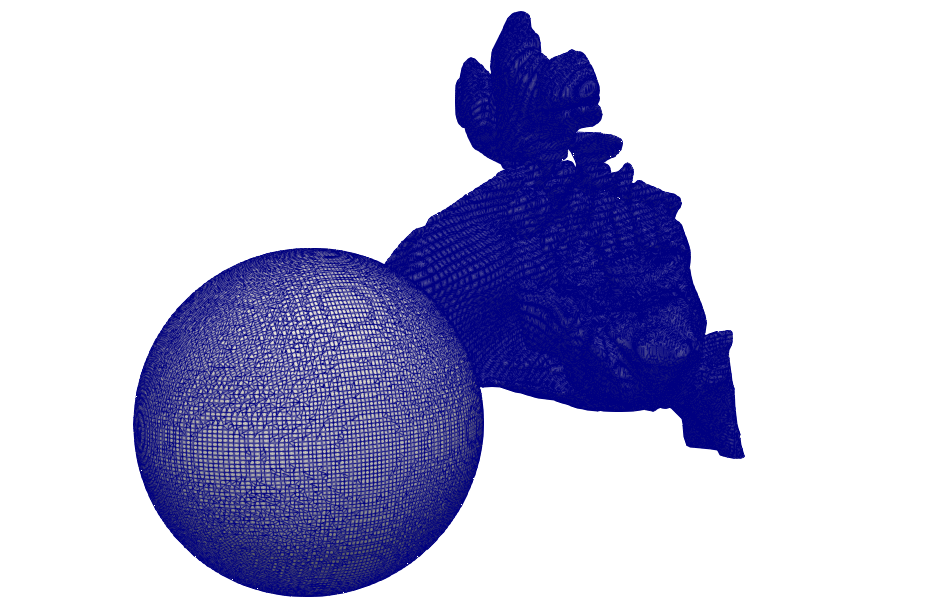}
\caption{Computational mesh for patient P2.}
\label{fig:P2-mesh}
\end{figure}

Figure \ref{fig:P2-mesh} shows the volume mesh obtained at the end of this procedure for P2. For a better description of the anatomy, finer cells are placed in correspondence of the nasal airways and of the paranasal sinuses, whereas a coarser grid is used for the inlet sphere. No layers are added, since the background mesh and the refining implicit in the adaptation process create fine enough cells near the boundaries. Table \ref{tab1} reports, for each of the three patients, the number of cells for the entire volume mesh and for the boundary of the nasal airways alone. The volume mesh is made by a number of cells ranging from 5.3 to 6.6 millions, whereas about 1 million cells describe the solid boundaries, i.e. the nasal walls. The small differences between meshes for the three patients are due to the different dimensions of the anatomies, since an identical refinement strategy was used. 
Overall, the CT scans are of standard quality, and the size of the computational meshes is typical for comparable RANS studies of the nasal airflow \cite{inthavong-etal-2019}.

\begin{table}
\centering
\caption{Number of cells in the volumetric mesh for the three patients.}
\label{tab1}%
\begin{tabular}{ccc}
\toprule
& \multicolumn{2}{c}{Number of cells} \\
& Volume Mesh & Solid Boundary\\
\midrule
\textbf{P1}   & 5262788 & 950327  \\
\textbf{P2}   & 5695738 & 1052087 \\
\textbf{P3}   & 6623922 & 1178278 \\
\bottomrule
\end{tabular}
\end{table}

\subsection{Direct RANS simulations}

The mathematical flow model of choice is the steady incompressible RANS equations, obtained after time-averaging the Navier--Stokes equations. It represents the most common choice for such problem, and accounts for turbulence via a turbulence model. 
The steady Navier--Stokes equations \eqref{eq:direct} written for the mean fields are augmented with the divergence of the apparent stress tensor $\mathbf{\overline{u'u'}}$ called tensor of the Reynolds stresses:
\begin{equation}
\begin{cases}
\nabla \cdot \mathbf{\overline{u}} = 0 \\
\nabla \cdot \left( \mathbf{\overline{u}} \, \mathbf{\overline{u}} \right) + \nabla \cdot \left( \mathbf{\overline{u'u'}} \right) = - \nabla \overline{p} +\nu \nabla^2 \mathbf{\overline{u}}
\end{cases}
\label{eq:RANS}
\end{equation}
where an overbar $\overline{\cdot}$ indicates the time-averaging operator, and a prime denotes fluctuations. These equations become closed and thus solvable only after a turbulence model specifies the functional form of the Reynolds stress tensor in terms of the mean velocity and its gradients.

The finite-volumes flow solver is OpenFOAM \cite{weller-etal-1998}. To carry out a meaningful multi-patient study, as discussed in Ref.\cite{segalerba-etal-2024}, we opt for a comparison at the same volumetric flow rate (CFR). A flow rate of $15 \ l/min$ is thus enforced, corresponding to a mild inspiration. A representative value of the Reynolds number for the three patients can be computed, at the outlet, interpreted as a straight non-circular duct. Defining the Reynolds number $Re$ based on the bulk velocity at the outlet, the hydraulic diameter and the kinematic viscosity of air, one obtains $Re=1727$ for P1, $Re=2054$ for P2 and $Re=1401$ for P3.
Our modeling choices are in agreement with relevant works that proved their accuracy at comparable flow rates, see e.g. \cite{liu-etal-2007, hoerschler-schroeder-meinke-2010}. The employed turbulence model is the $k-\omega$ SST turbulence model, chosen because of its ability to provide reasonable results while being numerically stable, and also because of its prevalence in the CFD studies of the nasal airflow \cite{islam-etal-2020}. Default values for the model coefficients are used. 
The differential system \eqref{eq:RANS} is provided with boundary conditions. At the inlet, the required value of the flow rate is prescribed for the velocity; no-slip and no-penetration conditions are applied at the solid boundaries representing the mucosal lining. At the outlet, located in the nasopharynx, a zero-gradient condition is applied to the velocity vector. Pressure has zero gradient at the inlet and on the solid boundaries of the nasal cavities, whereas at the outlet an arbitrary reference value $p_0=0$ for pressure is specified: in the incompressible setting, only pressure differences have dynamical meaning. As for the turbulent quantities, the required boundary conditions are summarized in Table \ref{tab:turb_bc}.

\begin{table}[ht]
\begin{minipage}{\columnwidth}
\centering
\caption{Boundary conditions for turbulent quantities: turbulent kinetic energy $k$, eddy viscosity $\nu_t$ and turbulence frequency $\omega$.}
\label{tab:turb_bc}%
\begin{tabular}{ccc}
\toprule
& Inlet & Outlet \\
\midrule
$k$   & $k=0.01$   & $\partial k / \partial \textbf{n} = 0$  \\ 
$\nu_t$    & $\partial \nu_t / \partial \textbf{n} = 0$   & $\partial \nu_t / \partial \textbf{n} = 0$  \\ 
$\omega$    & $\omega=0$   & $\partial \omega / \partial \textbf{n} = 0$ \\
\bottomrule
\end{tabular}
\end{minipage}
\end{table}

\subsection{Adjoint solution and surface sensitivity}

While the procedure outlined so far is standard, computing the adjoint solution and the surface sensitivity is way less common, and to our knowledge has never been attempted in the context of the nasal airflow. 

The system of the adjoint differential equations has been already presented in Sec.\ref{sec:theory}, where the PDE to compute the adjoint velocity $\mathbf{v}$ and the adjoint pressure $q$ were derived from the steady Navier--Stokes equations \eqref{eq:direct}. Here, we use the RANS equations \eqref{eq:RANS} as primal equations, hence the adjoint equations would have to include the effect of the turbulence model of choice.
However, we take advantage of the so called "frozen turbulence" assumption, according to which the variations of the turbulent quantities with respect to the control variables are negligible. Thus, the derivation of the adjoint counterparts of the complete RANS equations including the turbulence model is avoided. This reduces complexity and computational cost, while the negative consequences on the computed sensitivity map are negligible \cite{schramm-etal-2018}. 

Once the adjoint system is available, the cost function $f$ to drive the optimization must be chosen. This critical step is going to determine the boundary conditions for the adjoint problem. We choose the total dissipated power as the quantity to be minimized. Indeed, dissipation is linked to the resistance encountered by the nasal airflow, and clearly increases when obstructions are present. The dissipated power is written as the integral across the boundary of the net flux of mechanical energy, i.e. the sum of pressure and kinetic energy:  
\begin{equation}
f = \int_\Gamma \left( p+\frac{1}{2}u^2 \right) \mathbf{u} \cdot \mathbf{n} \ d \, \Gamma
\label{eq:goal}
\end{equation}

The dissipated power equals the viscous losses, but the equivalent expression above yields an easier expression to handle. Moreover, in this way $f$ explicitly depends on the flow variables only. Obviously, the control variables $\beta$, albeit not appearing in the cost function, are involved in the procedure through their role in the constraints $R$.

\begin{figure}[ht]%
\centering
\includegraphics[width=\columnwidth]{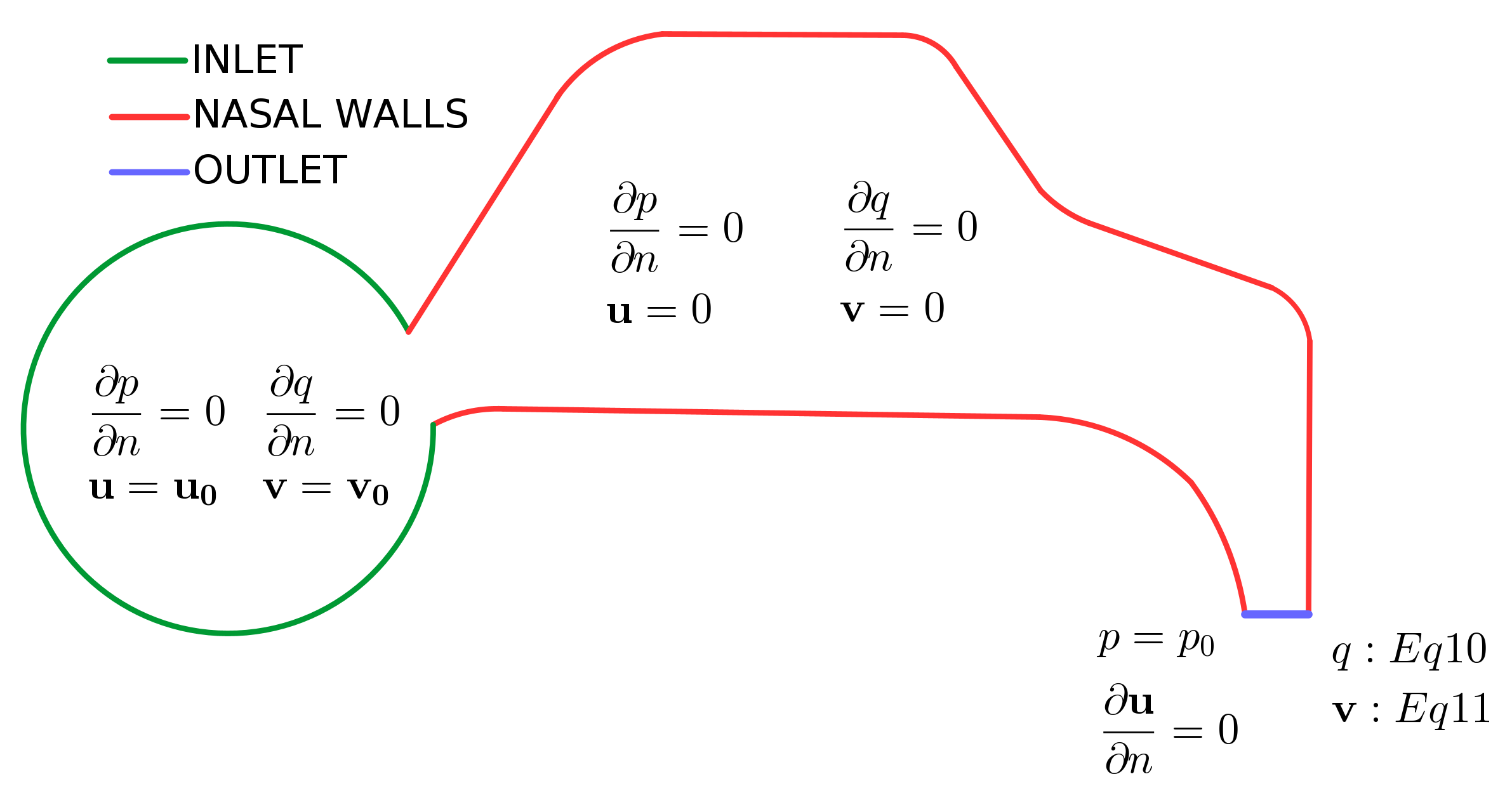}
\caption{Schematic representation of the computational domain with the boundary conditions for the direct ($\mathbf{u}$ and $p$) and adjoint ($\mathbf{v}$ and $q$) equations.}
\label{fig:bc}
\end{figure}

The definition of $f$ determines the boundary conditions for the adjoint equations. Again, details on their analytical derivation can be found e.g. in Ref.\cite{othmer-2008}: here we simply list them, and represent them schematically in figure \ref{fig:bc} together with those of the direct problem.  

At the outer ambient and at the solid walls, the conditions for $q$ and $\mathbf{v}$ are identical to those for $p$ and $\mathbf{u}$.  At the outflow boundary, instead, the derived boundary conditions are:
\begin{equation}
q = \mathbf{v} \cdot \mathbf{u} +v_n u_n + \nu \left( \mathbf{n}\cdot \nabla\right) v_n -\frac{1}{2} u^2 - u_n^2
\end{equation}
and
\begin{equation}
0=u_n \left( \mathbf{v}_t - \mathbf{u}_t \right) + 
\nu \left( \mathbf{n} \cdot \nabla \right) v_t 
\end{equation}
where the subscripts $n$ and $t$ refer to the component normal and tangential to the boundary. 

\section{Results}
\label{sec:results}

A brief and qualitative description of results from the standard direct RANS simulation is presented first in Sec.\ref{sec:direct}; the newly introduced adjoint fields are shown in Sec.\ref{sec:adjoint}. For brevity, only P2 is considered, as the one who presents the most evident anomalies. 
The sensitivity maps computed for P1, P2 and P3 are specifically addressed later in Sec.\ref{sec:sensitivity}. 

\subsection{RANS}
\label{sec:direct}

\begin{figure}[ht]%
\centering
\includegraphics[width=\columnwidth]{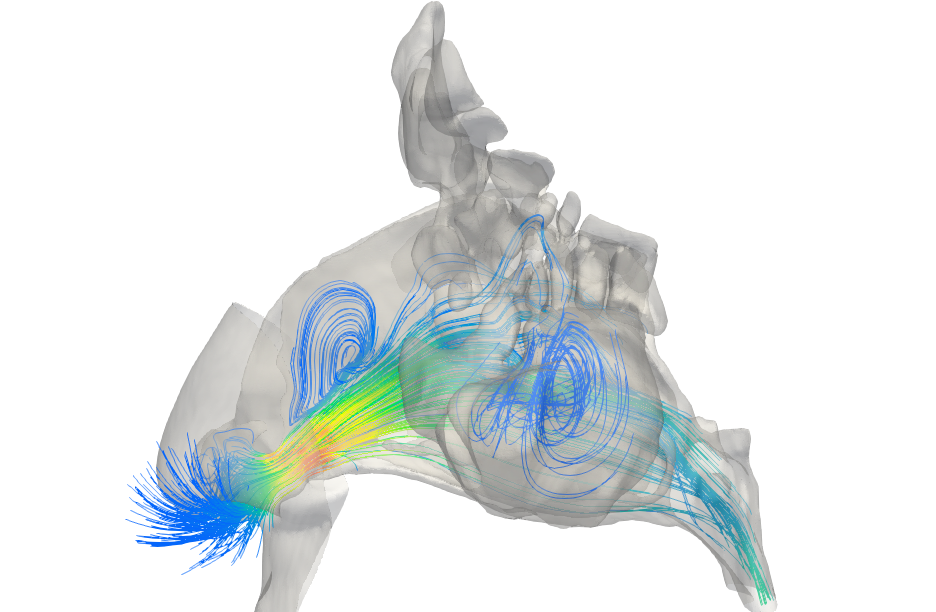}
\includegraphics[width=\columnwidth]{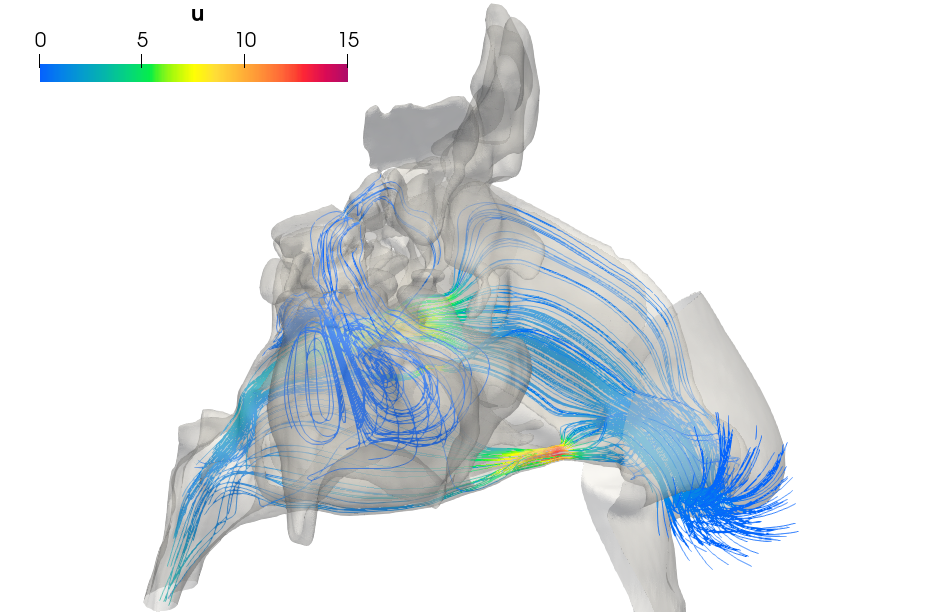}
\caption{Patient P2: mean streamlines in the left (top) and right (bottom) nasal cavity, colored by the magnitude of the mean velocity.}
\label{fig:streamlines-P2}
\end{figure}

Figure \ref{fig:streamlines-P2} shows two three-dimensional views of the mean streamlines for P2, for the left and right passageways. The flow undergoes first a significant acceleration near the nose tip on both sides from the nearly still external air to more than $2 \ m/s$. A large recirculation is then observed after the left deviation of the quadrangular cartilage. Here the velocity magnitude is around $1 \ m/s$, which represents its minimum in both nasal fossae. A further evident consequence of the septal deviations is the flow acceleration in the restriction of the inferior right meatus. Here a very large peak value of $13 \ m/s$  for the velocity magnitude is reached. Globally, obstructions may cause an asymmetry between the left and right nasal cavities; for P2, $9.33 \ l/min$ and $5.67 \ l/min$ are the left and right volumetric flow rates, corresponding to 62.2\% of the flow through the left passage and the remaining 37.8\% through the right one.

\begin{figure}[ht]%
\centering
\includegraphics[width=\columnwidth]{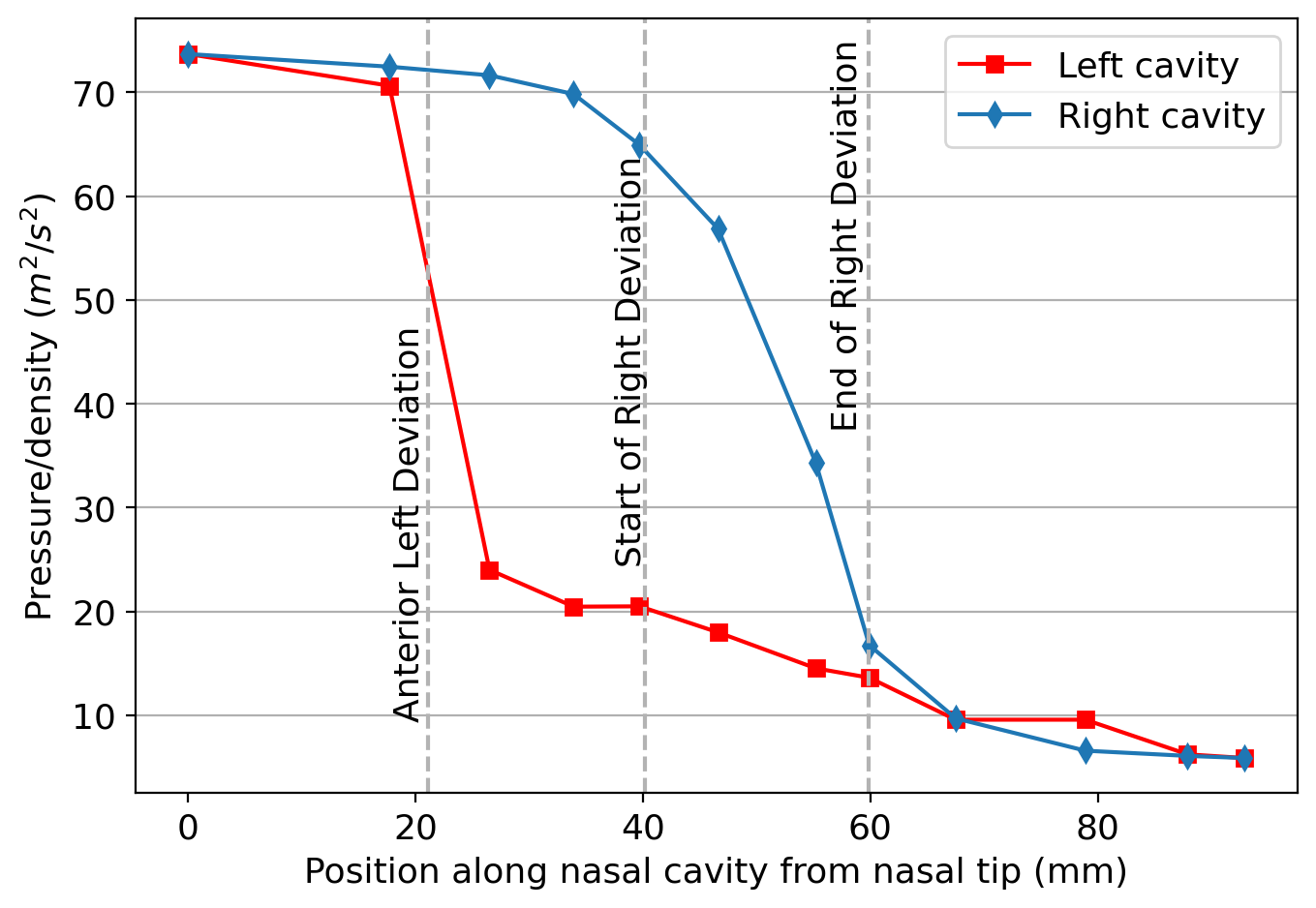}
\caption{Patient P2: evolution of the mean section-averaged pressure (divided by density) along the right and left nasal fossae, from the nose tip to the choana.}
\label{fig:pressure-P2}
\end{figure}

Figure \ref{fig:pressure-P2} quantifies the asymmetry between the two cavities in terms of the evolution of the mean section-averaged pressure from the nasal tip to the choanae, and helps identifying regions of local pressure losses. 
Pressure values (divided by density, as it is customary within OpenFOAM with incompressible flows) are computed at twelve locations, by averaging the mean pressure field over the entire local cross-sectional area. 
In both cavities, significant losses are present in correspondence of obstructions. In the left fossa, the anterior deviation causes a sudden and localized decrease from $70 \ m^2/s^2$ to $23 \ m^2/s^2$; after the anomaly, pressure decreases smoothly. In the right fossa, instead, although flow perturbations are even more severe at both points where the bone spur reaches the turbinates, the section-averaged losses appear to be milder; they are not concentrated at one specific section, but involve a rather large tract that extends for $25 \ mm$ along the axis of the fossa.

\subsection{Adjoint field}
\label{sec:adjoint}

\begin{figure}[ht]%
\centering
\includegraphics[width=\columnwidth]{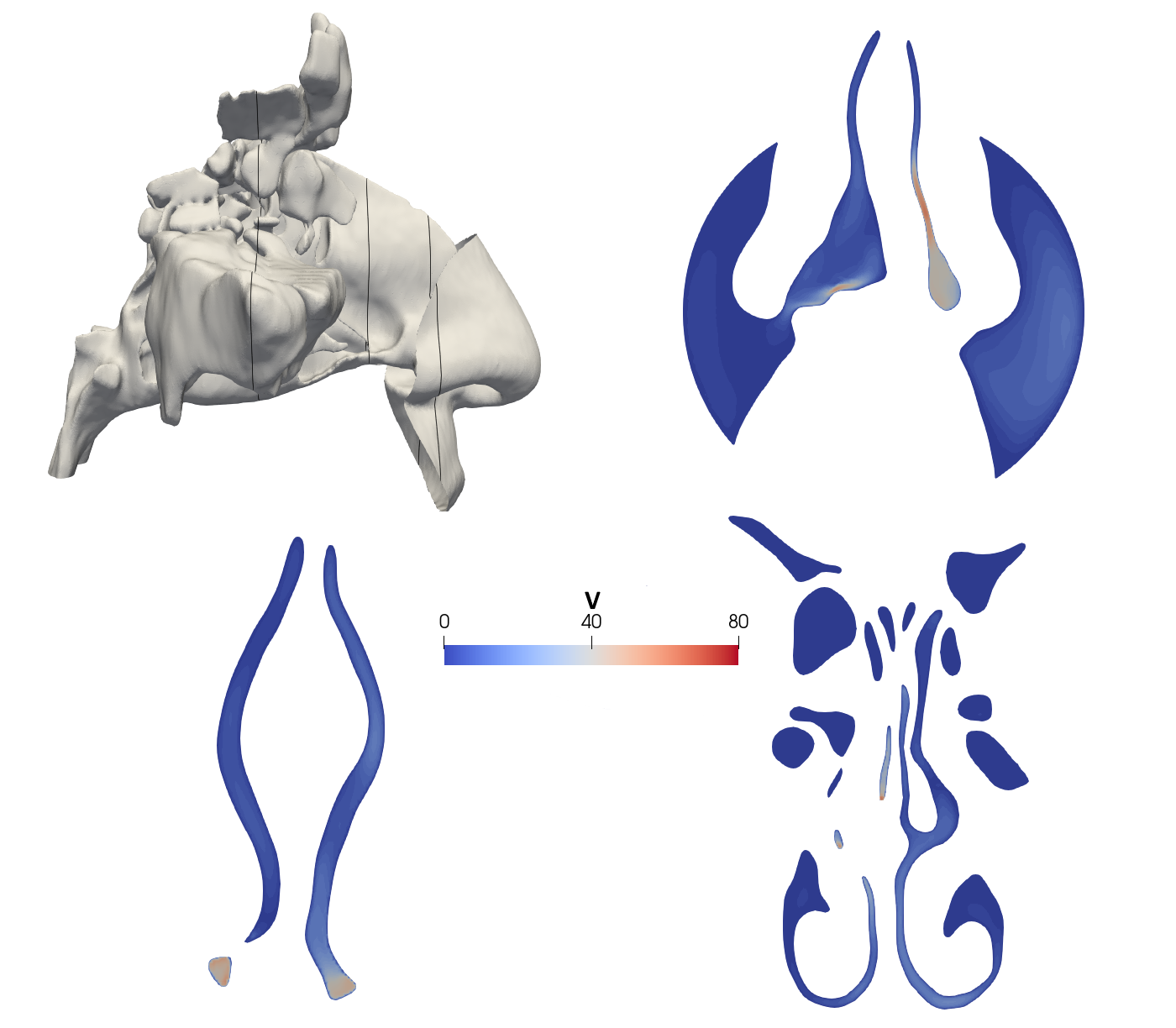}
\caption{Patient P2: adjoint velocity magnitude, visualized in the three sections highlighted in the top left panel, taken in correspondence of major obstructions.}
\label{fig:adjvelocity-P2}
\end{figure}

\begin{figure}[ht]%
\centering
\includegraphics[width=\columnwidth]{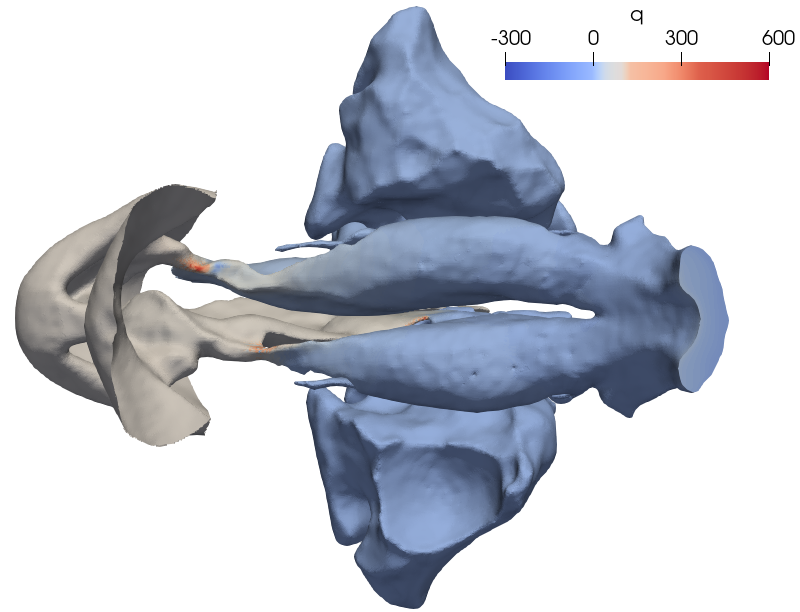}
\caption{Patient P2: adjoint pressure.}
\label{fig:adjpressure-P2}
\end{figure}

The adjoint velocity and pressure fields are visualized, once again for patient P2 only, in figures \ref{fig:adjvelocity-P2} and \ref{fig:adjpressure-P2}. The adjoint velocity $\mathbf{v}$ and the adjoint pressure $q$ are not lending themselves to an immediate physical interpretation. They are mathematically defined fields, which depend directly on the chosen objective function expressed by Equation \eqref{eq:goal}: changing $f$ would lead to computing different adjoint fields, as the adjoint equations are unchanged but their boundary conditions depend on $f$. Both fields are seen to assume their maxima in qualitative correspondence to obstructed regions, as a consequence of the choice of the dissipated power as the objective function. This implies that these regions may be important for the effective removal of obstruction. However, quantitative information for optimization is obtained only when the adjoint field is combined with the primal field, via the surface sensitivity \eqref{eq:sensitivity}. This is discussed, for the three cases, in the remainder of this Section.

\subsection{Surface sensitivity}
\label{sec:sensitivity}

Combining the information contained in the direct and adjoint solutions into the surface sensitivity via Equation \eqref{eq:sensitivity} allows one to quantify, for each point on the mucosa, the potential minimization of the cost function that can be achieved by surgery. 
Sensitivities, which contain the most clinically important information, are discussed below for all three patients. Owing to the linear nature of the adjoint problem, a normalized surface sensitivity (represented hereafter with the letter $\eta$) is visualized, i.e. the surface sensitivity of each cell is divided by the corresponding global maximum. 

To interpret results from a surgical point of view, it should be kept in mind that the red color (i.e. high positive sensitivity) labels regions where the optimization indicates reduction of the cost function via a normal displacement of the surface that enlarges the cavity, i.e. consistent with the typical surgical action: the sensitivity takes its sign from the projection onto the wall-normal direction, which is oriented from the fluid region outwards. 
Regions in gray, instead, is where sensitivity is small; here the benefits of surgery towards minimization of the cost function are limited. 
Lastly, blue-colored regions suggest a local reduction of the cross-sectional area, and imply surgical reconstruction. 

\subsubsection{P1}

\begin{figure}[ht]%
\centering
\includegraphics[width=\columnwidth]{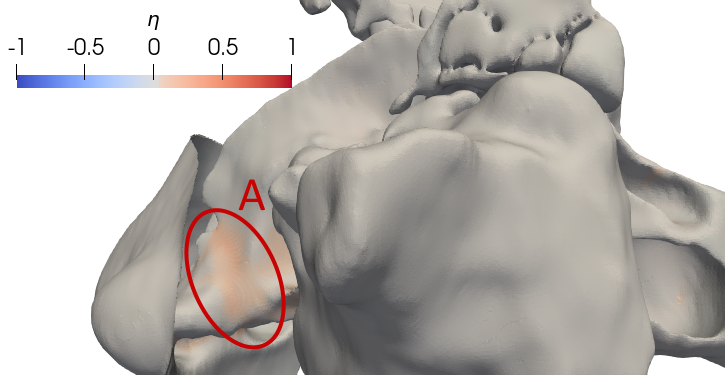}
\includegraphics[width=\columnwidth]{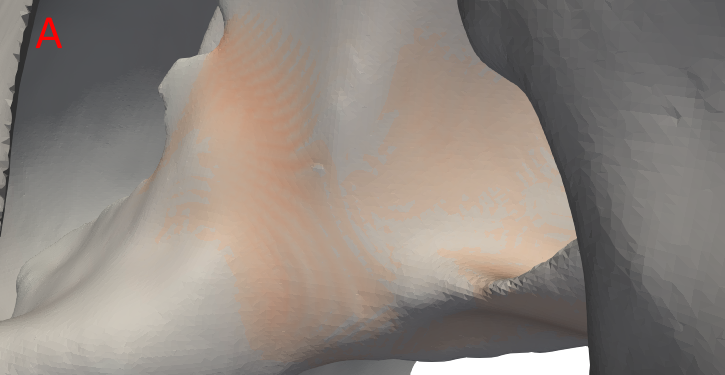}
\caption{Surface sensitivity for P1, region A. Top: lateral view of the left side. Bottom: zoom on the region marked by the red circle.}
\label{fig:P1a}
\end{figure}

\begin{figure}[t]%
\centering
\includegraphics[width=\columnwidth]{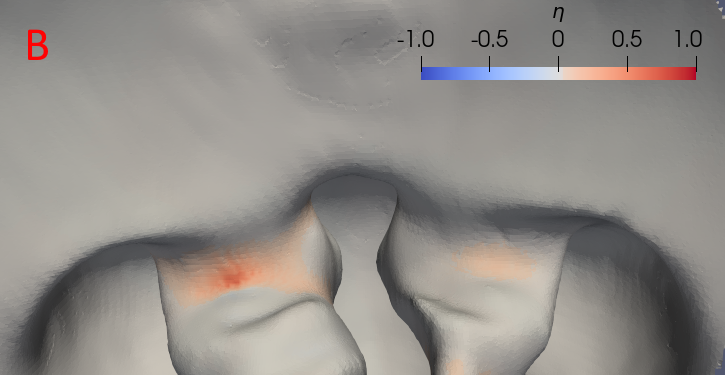}
\caption{Surface sensitivity for P1, region B. View from below.}
\label{fig:P1b}
\end{figure}

\begin{figure}[ht]%
\centering
\includegraphics[width=\columnwidth]{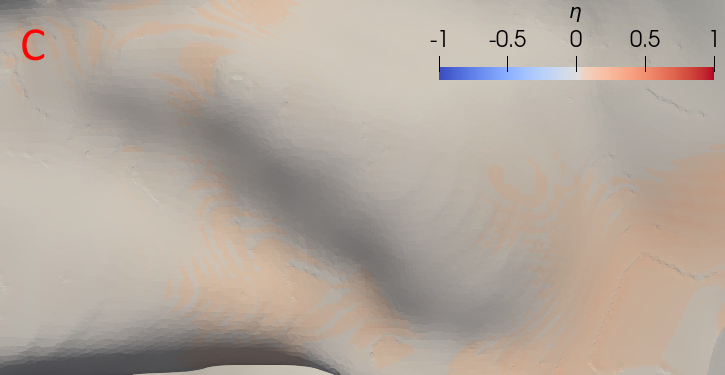}
\caption{Surface sensitivity for P1, region C. View from the right of the left nasal cavity in correspondence of the left deviation}
\label{fig:P1c}
\end{figure}

The surface sensitivity for patient P1 is shown in Figures \ref{fig:P1a}, \ref{fig:P1b} and \ref{fig:P1c}. For this patient, the adjoint optimization identifies three main areas with large sensitivity: region A in correspondence of the deviation of the quadrangular cartilage, region B behind the right nostril, and region C located along the left nasal fossa where the septum is deviated. 

Figure \ref{fig:P1a} shows a lateral view of the left nasal airway, where region A is located; below, a zoomed-in view highlights its characteristics. Local maxima for the sensitivity are found in correspondence of the obstruction, where cells have normalized values of about $0.2$. 
Similar values (above $0.1$) are also obtained for the lateral mucosal walls, thus pointing at a functional link between septoplasty and turbinoplasty.
Even though not visible, a similar red area (with lower sensitivity values) is also found on the same part of the nasal fossa but medially towards the septum. Hence, the adjoint procedure suggests an enlargement of the entire cross-sectional area.

To analyze region B, Figure \ref{fig:P1b} provides a view of the area behind the nostrils. This region hosts the cell with maximum sensitivity (normalized at unitary value) on the right nostril. Nearby, a relative large area with sensitivity values above 0.25 is found. Also, a few cells with very small negative sensitivities are observed.

Finally, region C is detected in correspondence of the left deviation of the vomer bone, and is shown in Figure \ref{fig:P1c}. Here, rather small normalized sensitivities around $0.03$ are obtained, suggesting a displacement that corrects the distortion. This deviation, evident when the CT scan is examined by an expert, is correctly identified by the adjoint, even though it does not cause a severe obstruction.

\subsubsection{P2}

\begin{figure}[ht]%
\centering
\includegraphics[width=\columnwidth]{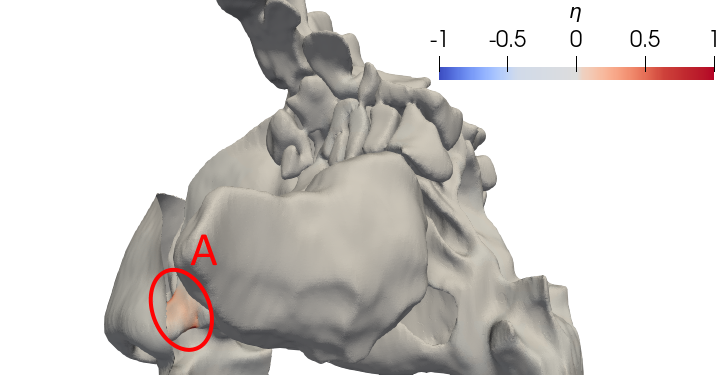}
\includegraphics[width=\columnwidth]{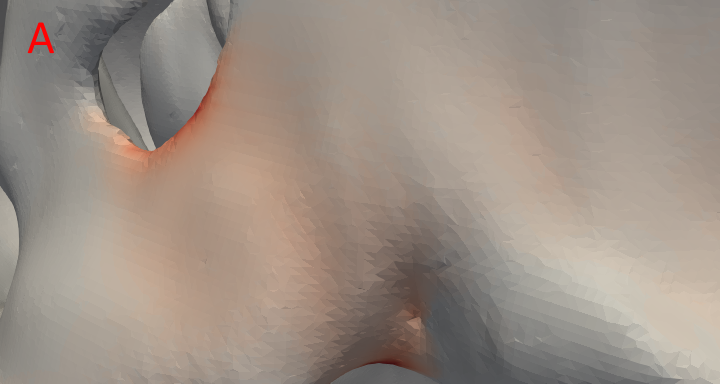}
\caption{Surface sensitivity for P2, region A. Top: lateral view of the left side. Bottom: zoom on the region marked by the red circle.}
\label{fig:P2a}
\end{figure}

\begin{figure}[t]%
\centering
\includegraphics[width=\columnwidth]{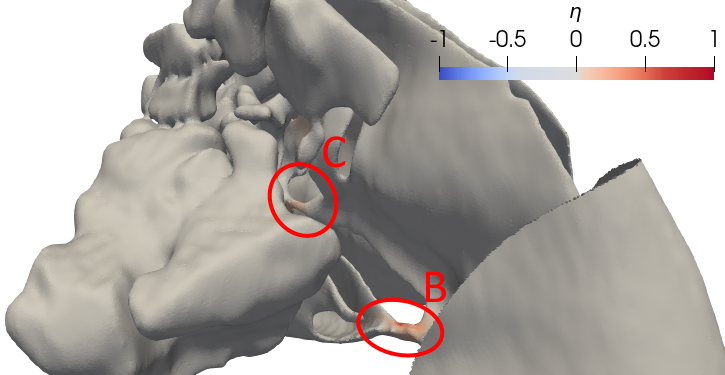}
\includegraphics[width=\columnwidth]{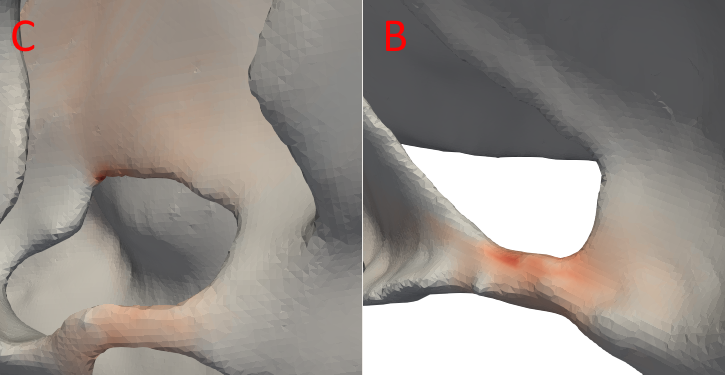}
\caption{Surface sensitivity for patient P2. Top: lateral view from the right. Bottom: zoom on the two regions emphasized by the red circles.}
\label{fig:P2b}
\end{figure}

The surface sensitivity for patient P2 is shown in Figures \ref{fig:P2a} and \ref{fig:P2b}. 

The first focuses on the sensitivity obtained in region A, corresponding to the left anterior deviation. The outcome of the adjoint here resembles that of region A for P1: the critical region is correctly identified, and the sensitivity suggests to surgically enlarge the entire cross-sectional area. 
However, some differences with P1 are observed when the local values of sensitivity are considered.
First of all, for P2 this area corresponds to the global maximum of sensitivity. Furthermore, the area showing large values of the sensitivity is wider, suggesting a more distributed surgical action.

Figure \ref{fig:P2b} shows a global view of the nasal cavities seen from the patient's right, and emphasizes two other critical regions, denoted as B and C, visible in more detail in the zoomed-in views. 
The right bone spur obstructs the flow by touching the inferior turbinate in region B, and the middle one in region C. 
In both cases, the adjoint procedure suggests to enlarge the restriction, but larger sensitivities are seen in correspondence of the inferior meatus, where sensitivity values of about $0.4$ are found in the top part of the restriction; in other points of the same restriction the sensitivity is about $0.2$. For the middle turbinate restriction, the adjoint optimization suggests to act on two locations. 
One corresponds to the middle meatus, where all cells have comparable sensitivity values around $0.1$. The second location, instead, is the part of the left nasal cavity located between the deviated septum and the turbinate, and is characterized by normalized values around $0.04$.

\subsubsection{P3}

\begin{figure}[ht]
\centering
\includegraphics[width=\columnwidth]{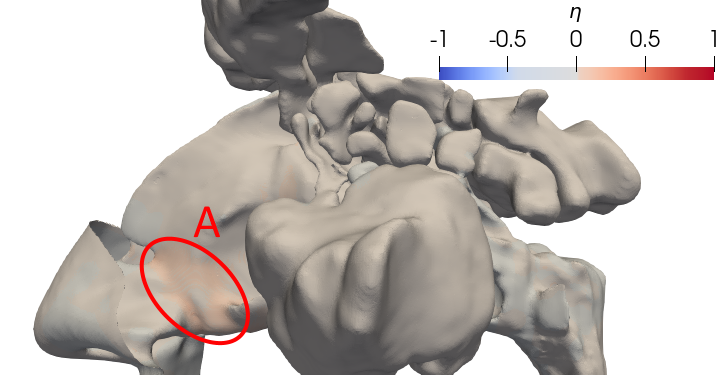}
\includegraphics[width=\columnwidth]{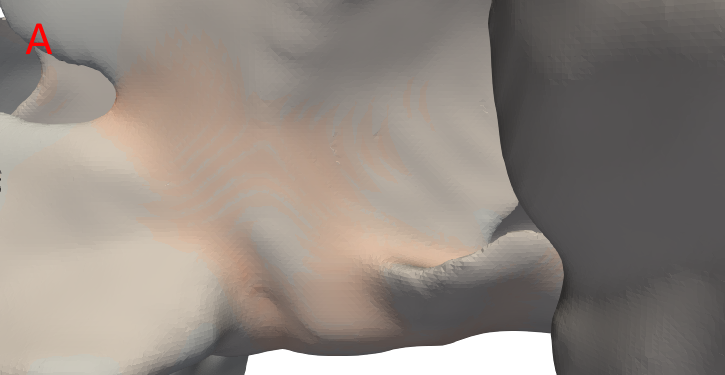}
\caption{Surface sensitivity for patient P3, region A. Top: left lateral view. Bottom: zoom on the region marked by the red circle.}
\label{fig:P3a}
\end{figure}

\begin{figure}[t]%
\centering
\includegraphics[width=\columnwidth]{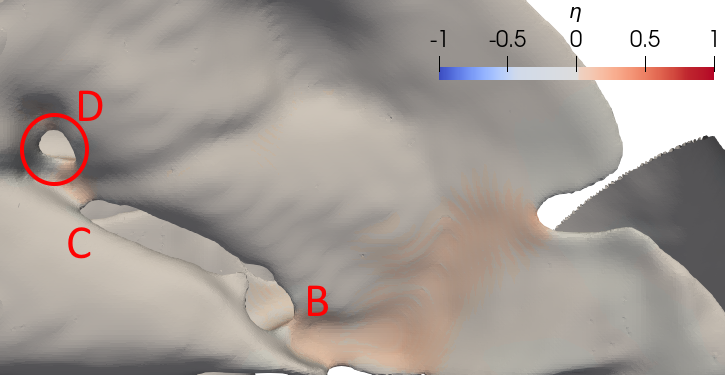}
\includegraphics[width=\columnwidth]{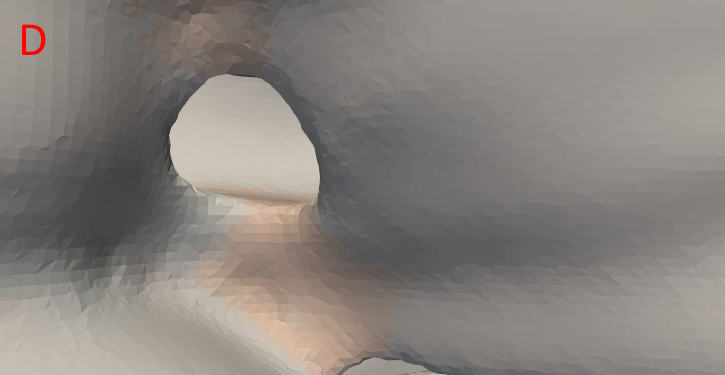}
\caption{Surface sensitivity for patient P3, regions B, C and D. Top: view of the left nasal cavity from the "inside". Bottom: zoom on the region emphasized by the red circle.}
\label{fig:P3b}
\end{figure}

\begin{figure}[ht]%
\centering
\includegraphics[width=\columnwidth]{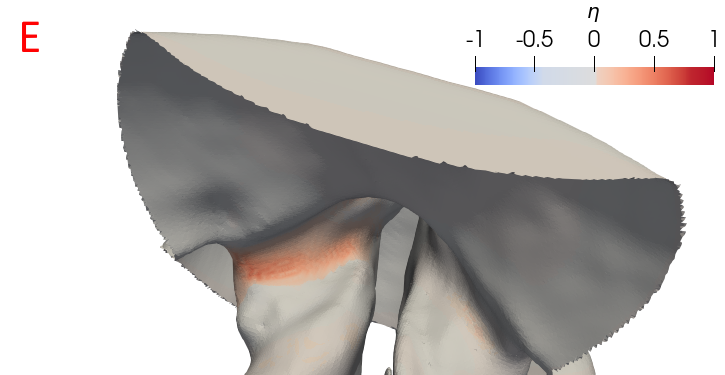}
\caption{Surface sensitivity for patient P3, region E.}
\label{fig:P3c}
\end{figure}

Figures \ref{fig:P3a}, \ref{fig:P3b} and \ref{fig:P3c} show the surface sensitivity computed for patient P3. For this patient, five critical regions are identified. 
Region A is the obstruction caused by the collapse of the nasal valve; regions B and C are the start and end points of the contact region between the left septal deviation and the inferior turbinate; region D is the obstruction caused by the left condro-vomerian spur that reaches the middle turbinate; region E is behind the right nostril.

Figure \ref{fig:P3a} plots region A. Here, the optimization suggests to operate with an outward displacement on the entire cross-section where the nasal valve is collapsed. The highest sensitivities are found in the top and bottom parts of the section, and are about $0.1$, whereas the surrounding red cells are about $0.04$. 
A corresponding red area is also found in the internal part of the nasal cavity, with a sensitivity around $0.3$ as shown in Figure \ref{fig:P3b}, which is a view on the left nasal cavity as seen from the inside. This visualization, obtained via clipping of the geometry, highlights also the critical regions B and C caused by the left septal deviation. 
First, the starting point of obstruction with the inferior turbinate can be analyzed. Here the solver computes values of the normalized sensitivity around $0.04$ and this red region also involves a small portion of the meatus and the bottom part of the nasal cavity. In correspondence of the end of the contact, instead, smaller values around $0.025$ are observed. Region D is highlighted in the zoomed-in view. Here the adjoint procedure gives a normalized value of $0.015$.

Lastly, Figure \ref{fig:P3c} shows region E, where sensitivities around  $0.25$ are found, with some cells peaking at about $0.35$.

\section{Discussion}
\label{sec:discussion}

We have introduced a novel approach to septoplasty planning, which has the potential to yield an objective method for surgery planning. The procedure employs relatively standard tools for the execution of patient-specific CFD simulations of the nasal airflow, and then leverages adjoint optimization, a technique applied for the first time in the present context.
The former (direct) part of the procedure, which includes the segmentation of the CT scan, the creation of the computational mesh, and the set up of the CFD simulation, does not require special considerations here. Although several important and critical steps are involved in that part, they are not discussed in this paper, since abundant literature is available. We note, however, that the robustness of the direct procedure with respect to those steps carries forward to the adjoint part too.

The original part of the procedure is the application of the adjoint-based optimization, which indicates where, along the mucosal wall, a surgical action can be most effective. 
The critical sections along the airways, often but not always corresponding to local minima of the cross-sectional area, are identified: moreover, the solution highlights where along the contour of the section(s) the surgical modifications are most effective.
This clinically important indication derives from the ability of the adjoint optimization to combine anatomical and functional information, the latter descending from the solution of the direct problem: in the end, the large sensitivities identified by the optimization arise where there is an obstruction to the flow, accompanied by a potentially much larger flow once said obstruction is removed. 
When the consequences of surgery are judged by only considering anatomy, as surgeons traditionally and necessarily do, the last factor is entirely missing. Three-dimensional maps of normalized sensitivity are immediately interpretable by an ENT surgeon: she can appreciate at a glance where it is suggested to act, and the details of how the obstruction should be reduced to maximize the benefits and minimize the invasiveness of the procedure.

The optimization procedure has been preliminarily applied to three selected cases of patients presenting complex and diverse septal deviations. The adjoint formulation correctly highlights every portion of the nasal septum that was previously identified during a pre-operative inspection of the CT scans carried out by experienced ENT surgeons. 
The computed sensitivity is very low over most of the turbinates, paranasal sinuses and at the nasopharynx, in agreement with the clinical observation that none of the patients presents significant turbinate hypertrophy. 
The informed opinion of the ENT surgeons is that the surgical actions suggested by the adjoint are deemed surgically feasible, with the section being locally enlarged through the displacement of the septal wall. The only exception is for patient P3 (see figure \ref{fig:P1a}), where in region A a displacement of the collapsed nasal valve was suggested. The collapse of the nasal valve is a frequent occurrence caused by fluid-structure interaction e.g. during intense inspiration or sniffing, but is not usually observed for normal breathing, and is not surgically corrected. Hence, although enlarging the nostrils is definitely meaningful in purely fluid mechanical terms, the sensitivity map obtained around the nasal valve should be considered with care or not considered altogether. 

To make the optimization procedure clinically viable, it is important for it to be as streamlined as possible, reducing the required operator time to a minimum. 
Presently, the segmentation of the CT scan is the only step that requires external supervision to check the quality of the reconstruction. As such, it is the most expensive part in terms of operator time, and it requires 10-20 minutes, depending on the expertise of the operator. 
The steps after segmentation can be easily automated, as they can be set up once and for all, independently from the specific anatomy.
In terms of computing time, on a conventional laptop it takes about 10 hours to compute the surface sensitivity map starting from the reconstructed geometry. More performing hardware and parallel computing can reduce these figures almost at will. Moreover, the flow model can be simplified further to arrive at equivalent results in a shorter time.
Interpretation of results might be facilitated by computing sensitivities only for subset(s) of the surface nodes where surgery is possible and meaningful. 

A noteworthy point concerns the robustness of the results. 
Computing sensitivities would certainly be affected positively by an high-resolution CT scan, but the sensitivities discussed above have been obtained by employing CT scans of standard quality and different spatial resolution: the axial spacing ranges from a rather low $0.925 \ mm$ for P1 and P2 to an average $0.4 \ mm$ for P3, thus including most -- if not all -- the routinely acquired scans: adjoint optimization can be applied to ordinary scans without the need of a dedicated protocol.

The outcome of the optimization is also robust with respect to the discretization of the numerical simulations. 
This has been assessed by re-running all cases on a coarser mesh, where the total number of cells was nearly halved (3.06 millions for P1, 3.17 millions for P2, and 4.1 millions for P3), finding that the indications provided by the adjoint optimization and by the wall sensitivity remain unchanged. That the optimization is robust with respect to (reasonable) variations of the spatial discretization is by no means an obvious result, especially within the present surface-based formulation of the adjoint problem.

Also, the outcome of the optimization is robust with respect to the choice of the turbulence model. Simulations were repeated (for P3 only) by using two different models, namely the $k-\epsilon$ model and the $v^2-f$ RANS model, in addition to the $k-\omega$ SST model. Results compared in terms of normalized sensitivity turn out to be unchanged within a relative 1\% in every cell of the domain. 
However, it should be stressed that the adjoint approach presented in this work is tightly connected to the use of the RANS approach. Although RANS is the most popular approach, and the nasal flow in the surgically significant areas is often steady or nearly so, using RANS for the nose flow, especially at low flow rates below $15 \ l/s$, is questionable and has been questioned \cite{hoerschler-schroeder-meinke-2010, naftali-etal-2005, calmet-etal-2021}.

Besides computational cost, the procedure as it stands presents additional features that might make its deployment difficult in a clinical setting. However, all such limitations can be properly addressed. For example, in the current implementation the adjoint sensitivity is computed over the whole computational boundary, but it can be limited to predetermined portions of the anatomy, and/or to only consider positive sensitivities, corresponding to tissue removal. Moreover, both the numerical code that computes the sensitivity (here, OpenFOAM), the equations of motion for which the adjoint problem is formulated (here, the RANS equations closed with the $k-\omega$ SST model and the frozen-turbulence hypothesis) and the adjoint formulation itself (here, based on surface sensitivity) are not unique or final: work is in progress on each of these fronts. As discussed below in Sec.\ref{sec:cost-function}, the best cost function remains to be established. While the present one possesses physical meaning and works well for septal deviations, more complex options, including for example heat transfer characteristics, should be explored to deal with a more general class of problems.

\subsection{Validation by virtual surgery}

The adjoint computes a local optimum, and an iterative process is needed to yield the truly optimal solution for a given problem. This implies eventually large changes in shape, and requires a constrained optimization process; however, properly writing the constraints into the cost function is possibly more delicate than determining the right cost function itself. Luckily, however, in the present context the initial geometry, which is known to correspond to the pre-operative condition, is by definition the right geometry to start from. By seeking a minimally invasive surgery to restore the nasal function starting from the pre-op anatomy, the local optimum is exactly the solution of interest for the surgeon.

For a continuous adjoint optimization, validating the solution is rather challenging, as it would require comparing, at each control point, the computed sensitivity with the finite difference of results from two simulations in which that control point is altered. Since the large number of control points makes this approach not viable in practice, the solution can be verified to provide an improvement of the cost function. 
To provide such a verification here, we use virtual surgery to assess the computed sensitivities. The original or pre-op anatomies are thus modified according to the indications of the adjoint procedure, and post-op anatomies are created. The direct problem is then solved again on the post-op anatomies. Virtual surgery is carried out by morphing the pre-op anatomies, i.e. every point of the nasal walls is displaced in the wall-normal direction according to the local value of the normalized sensitivity. An arbitrary scaling factor of $1 \ mm$ is used to scale the maximum normalized sensitivity. Such displacements are quite small; the maximum value is chosen to guarantee a balance between the possibility of measuring an improvement and the need of controlling the quality of the final STL file automatically generated with morphing. 

Morphing of the anatomy is carried out by resorting to the mesh motion solver within OpenFOAM, where the boundary conditions for the displacement of each point of the STL surface are specified. The new STL files generated with such procedure are then compared with the original ones, and the area of contact with the inlet sphere is carefully evaluated to ensure that the morphing process does not lead to topological changes of the surface. 
The average displacement of the points for the three patients are $30 \ \mu m$, $39 \ \mu m$ and $41 \ \mu m$ for P1, P2 and P3 respectively.

\begin{table}[ht]
\begin{minipage}{\columnwidth}
\centering
\caption{Flow partitioning and nasal resistance computed for all three patients after optimization}
\label{tab:prepost}%
\begin{tabular}{ccccccc}
\toprule
&\multicolumn{3}{c} {Before} & After \\
& L [\%] & R [\%] & $R_{nose}$ & L [\%] & R [\%] & $R_{nose}$\\
\midrule
\textbf{P1} & 32.5 & 67.5  & 0.091 & 32.2 & 67.8 & 0.081 \\
\textbf{P2} & 62.2 & 37.8  & 0.360 & 60.6 & 39.4 & 0.311 \\
\textbf{P3} & 22.9 & 77.1  & 0.037 & 23.5 & 76.5 & 0.033 \\
\bottomrule
\end{tabular}
\end{minipage}
\end{table}

Table \ref{tab:prepost} reports, for all the pre-op and post-op anatomies, the computed values of flow partitioning, and the corresponding nasal resistance, defined as $R_{nose} = \Delta p / Q$, where $Q$ is the volumetric flow rate expressed in $ml/s$, and $\Delta p$ is the pressure drop, expressed in $Pa$, from the outer ambient to the nasopharynx. 
Different types of obstructions lead to a rather wide range of nasal resistances; flow partitioning too goes from highly asymmetrical (P3) to nearly normal (P2). In fact, none of these quantities is a general and robust indicator for NAO. 

The post-op anatomies indicate a rebalancing of the quantity of air passing through the two nasal fossae for P2 and P3; for P1, instead, a minor (less than 0.3\%) deterioration of the flow symmetry is observed.
At least at the first step of the optimization procedure, this outcome is certainly possible, as the objective function does not target flow symmetry directly. Hence, it constitutes a further indirect confirmation of the weak link between flow partitioning and NAO.

The nasal resistance, though, does decrease in all cases. Since the inlet flow rate is not changed, these variations imply a reduction of the pressure difference between the nostrils and the nasopharynx. Changes in the nasal resistance are quite significant, in the order of 5-10\%, to be evaluated against the very small maximum displacement, set at $1 \ mm$, which corresponds to a typical or averaged displacement, on the virtually operated areas, of less than $40 \ \mu m$. Although one cannot expect such a precision surgery to be realistically possible, these results indicate that the surgery suggested by the adjoint achieves large improvements of the nasal resistance with minuscule, minimally invasive corrections.

The minimalistic nature of the suggested surgical action, combined with the extreme effectiveness at improving the cost function, is indicative of the potential of the adjoint optimization at identifying those small surgeries that could just do the job. It is not unreasonable to expect that such rationally derived and minimally invasive surgeries should contribute to improving the success rate of septoplasties, as well as to ensuring a rapid post-operative recovery.

\subsection{The choice of the cost function}
\label{sec:cost-function}

Selecting the cost function is the most delicate step of the procedure, and one that is not entirely free from empiricism. 
In this work, using the dissipated power in Equation \eqref{eq:goal} is found to yield satisfactory results, with the identification and the characterization of all major anomalies reported by ENT doctors. 
However, this specific cost function is not expected to easily generalize to other obstructive problems; even with septal deviations, it remains to be assessed whether it represents the best possible option.
Alternative cost functions that could describe additional functions of the nasal airways should be considered in future analyses. The physiology of the nose is complex, and its many functions should be properly evaluated. 
In this regard, the linearity of the adjoint equations might be exploited to define a more general cost function, written as the sum of multiple contributions to represent the different feelings and needs of the patients. This would also require the redefinition of both the direct and the adjoint equations, accounting e.g. for heat exchange and humidification. Work is ongoing to make the optimization procedure fully aware of the rich physics of the airflow in the nose.

\section{Conclusions}
\label{sec:conclusions}

Septoplasty is known to be sometimes ineffective at relieving patients from symptoms induced by a deviated nasal septum. 
This is related, at least in part, to the lack of standardized patient-specific tools to evaluate each septal deviation, and to provide the surgeon with functional information.

In this study, we have introduced a CFD procedure that augments the conventional numerical study of the nasal airflow with an adjoint-based shape optimization, thus obtaining an effective tool for surgery planning.
Adjoint-based optimization is used to compute sensitivity derivatives with respect to a cost function that expresses the dissipated power and indirectly accounts for the nasal resistance.
The procedure naturally outputs the quantitative information needed by surgeons to decide where their efforts should be preferentially placed, and allows minimally invasive surgeries.

The validity of the adjoint-based procedure has been proved by applying it to three nasal anatomies affected by complex septal deviations for which the surgical correction is not self-evident. 
Inspection of the computed normalized surface sensitivity maps has demonstrated the ability of the method to automatically identify all the functionally important anatomical alterations. Furthermore, the surgical indications provided by the sensitivity have been validated by ENT surgeons. 
The robustness of the procedure with respect to several aspects of the computational procedure has also been proved.

Further progress is certainly needed for this CFD method to become a clinical tool that ENT surgeons can use in their daily practice. 
The formulation itself in terms of surface sensitivity has alternatives, and work is underway to assess what is the best approach for the specific optimization problem. 
Moreover, a cost function of more general validity should be conceived, depending on the generality one intends to achieve.
However, the present work represents a significant step towards a robust and patient-specific approach for computer-assisted septoplasty planning.




\bibliographystyle{cas-model2-names}
\bibliography{../Nose}



\end{document}